\documentstyle[multicol,aps,epsf]{revtex}
\begin{document}
\title{Dynamics of Vibrated Granular Monolayers}
\author{X.~Nie$^{1}$, E.~Ben-Naim$^{1}$, and S.~Y.~Chen$^{1,2}$}
\address{${}^{1}$Theoretical Division and Center for Nonlinear Studies,
Los Alamos National Laboratory, Los Alamos, NM 87545\\
${}^{2}$IBM Research Division, T. J. Watson Research Center,
P.O. Box 218, Yorktown Heights, NY 10598}
\maketitle

\begin{abstract}
  
  We study statistical properties of vibrated granular monolayers
  using molecular dynamics simulations. We show that at high
  excitation strengths, the system is in a gas state, particle motion
  is isotropic, and the velocity distributions are Gaussian. As the
  vibration strength is lowered the system's dimensionality is reduced
  from three to two.  Below a critical excitation strength, a
  gas-cluster phase occurs, and the velocity distribution becomes
  bimodal. In this phase, the system consists of clusters of immobile
  particles arranged in close-packed hexagonal arrays, and gas
  particles whose energy equals the first excited state of an isolated
  particle on a vibrated plate.

\end{abstract}
\begin{multicols}{2}
  
  Granular media, i.e, ensembles of hard macroscopic particles exhibit
  rich, interesting, and only partially understood collective behavior
  \cite{jnb,lpk,dg}.  The dynamics of driven or excited granular media
  is particularly important since the external energy source balances
  the energy loss due to collisions. Collective behavior of such
  systems is therefore important for establishing a more complete
  theoretical description of granular media. Here, we focus on
  monolayer geometries where collapse, clustering, and long range
  order have been observed recently
  \cite{olafsen1,olafsen2,losert1,losert2,dlk,gzb}. In particular,
  experimental studies reported that the velocity distribution
  function may exhibit both Gaussian and non-Gaussian behavior under
  different driving conditions.
    
  In this study, we carry out molecular dynamics simulations of
  vertically vibrated granular monolayers. To validate the simulation
  method, we verified the experimentally observed transition from a
  gas-like phase in high vibration strengths, to a cluster-gas phase
  in low vibration strengths \cite{olafsen1,olafsen2,losert2}.
  Additionally, we checked that several other details including the
  transition point, and the statistics of the horizontal velocities
  agree quantitatively with the experimental results. In particular,
  the horizontal energy vanishes linearly near the transition point,
  and the corresponding velocity distribution changes from a Gaussian
  to a non-Gaussian as the vibration strength is reduced.
  
  In contrast with the experimental studies, the simulations enables
  us to probe the vertical motion, an important characteristic of the
  dynamics. Our results show that as the system approaches the
  transition point, the vertical energy drops by several orders of
  magnitude. Furthermore, the gas-cluster phase is characterized by a
  coexistence of clusters of immobile particles, and energetic gas
  particles, whose energy can be understood by considering an isolated
  particle on a vibrated plate. We also find that the deviation from
  the Gaussian behavior in the gas phase is directly related to the
  development of an anisotropy in the motion, i.e, significant
  differences between the horizontal and the vertical velocities.
  
  To study the dynamics of vibrated monolayers, we used the standard
  molecular dynamics simulation technique \cite{herrmann}. We
  considered an ensemble of $N$ identical weakly deformable spheres of
  mass $m$, radius $R$, and moment of inertia $I\cong {2\over 5}mR^2$.
  The simulation integrates the equations of motion for the linear and
  angular momentums, $m\ddot{\bf r}_i = \sum_{j\ne i}{\bf
    F}^n_{ij}+mg\hat z$, and $I\dot {\bf \omega}_i=\sum_{j\ne i}{\bf
    r}_{ij}\times {\bf F}^t_{ij}$, respectively.  Here, ${\bf r}_i$ is
  the position of the $i$th particle, $\omega_i$ is its angular
  velocity and $g$ is the gravitational acceleration. The force due to
  contact with the $j$th particle in the direction normal (tangential)
  to the vector ${\bf r}_{ij}={\bf r}_j-{\bf r}_i$ is denoted by ${\bf
    F}^n_{ij}$ (${\bf F}^t_{ij}$).  The force between two particles is
  nonzero only when they overlap, i.e., ${\bf F}_{ij}=0$ when $|{\bf
    r}_{ij}|>2R$.  When there is an overlap, the normal contact force
  ${\bf F}^n_{ij}={\bf F}^{\rm rest}_{ij}+{\bf F}^{\rm diss}_{ij}$
  between the particles is a sum of the following forces: (a) A
  restoring force, ${\bf F}^{\rm rest}_{ij} = Y m_i(|{\bf r}_{ij}| -
  2R) {\bf r}_{ij}/|{\bf r}_{ij}|$, with $Y$ the Young's modulus, and
  (b) An inelastic dissipative force, ${\bf F}^{\rm diss}_{ij} = -
  \gamma_n m_i{\bf v}_{ij}^{n}$.  The tangential force is the
  frictional force ${\bf F}^t_{ij}={\bf F}^{\rm shear}_{ij}= -\gamma_s
  m_i{\bf v}_{ij}^{t}$.  In the above, ${\bf v}_{ij}^{n}=({\bf
    v}_{ij}\cdot{\bf r}_{ij}){\bf r}_{ij}/|{\bf r}_{ij}|^2$, and ${\bf
    v}_{ij}^{t} = {\bf v}_{ij} - {\bf v}_{ij}^n $ are projections of
  the relative velocities in the normal and tangential directions,
  respectively. The coefficients $\gamma_n$ and $\gamma_s$ account for
  the dissipation due to the relative motion in the normal and
  tangential directions, respectively. Overall, the molecular dynamics
  method has the advantage that it is amenable for parallel
  implementation, and that it allows handling of collisions involving
  an arbitrary number of particles.
  
  Initially, particles are randomly distributed on the vibrating
  plate, with a filling fraction $\rho$.  The velocities were drawn
  independently from an isotropic Gaussian distribution.  The plate
  undergoes harmonic oscillations in the vertical direction according 
  to $z_p(t)=A(t) \cos(\omega t)$ with $A$ the vibration amplitude, and
  $\omega=2\pi\nu$ with $\nu$ the frequency.  When a particle collides
  with the plate, it experiences the same force as if it were to
  collide with another particle moving with the plate velocity. The
  simulation was carried in a finite box with a height chosen to be
  large enough so that no collisions can occur with the box ceiling.
  Periodic boundary conditions were implemented horizontally.
  Overall, the simulation parameters were chosen to be as compatible
  as possible to the experimental values \cite{olafsen1}: $R=0.595mm$,
  $g=9.8m/s^2$, $Y=10^7 /s^2$, $\nu=70Hz$, $N=2000$, $\rho=0.463$,
  $\gamma_s=100/s$, and $\gamma_n=200/s$. The above normal
  dissipation parameter leads to a restitution coefficient of
  $r=0.95$. Without loss of generality, we set the particle mass to
  unity, $m=1$.  We verified that the results reported in this paper
  were independent of the value of most of these parameters, as well
  as the nature of boundary conditions.

  We are primarily interested in statistical properties of the system
  in the steady state, and especially their dependence on the
  vibration strength, which can be quantified by the dimensionless
  acceleration $\Gamma = A \omega^2 /g $. The quantity $\Gamma$ can be
  tuned by varying either $\omega$ or $A$. We chose to fix the
  frequency and vary the vibration amplitude. This method should be
  valid as long as the time scale underlying the variation is larger
  than the systems' intrinsic relaxation time scales. Each of our
  simulations was initially run at $\sim 10^3$ oscillation cycles at a
  constant amplitude $A_0$. Then the amplitude was slowly
  reduced in a linear fashion according to $A/A_0 = 1-t/\tau$, with
  the decay time $\tau\cong 10^3 s$ (or alternatively $\sim 10^5$
  cycles). Throughout this paper we report measurements of average
  quantities such as the temperature and the velocity distribution.
  These were obtained by averaging over 100 consecutive oscillation
  cycles. 

\begin{figure}
\centerline{\epsfxsize=9cm \epsfbox{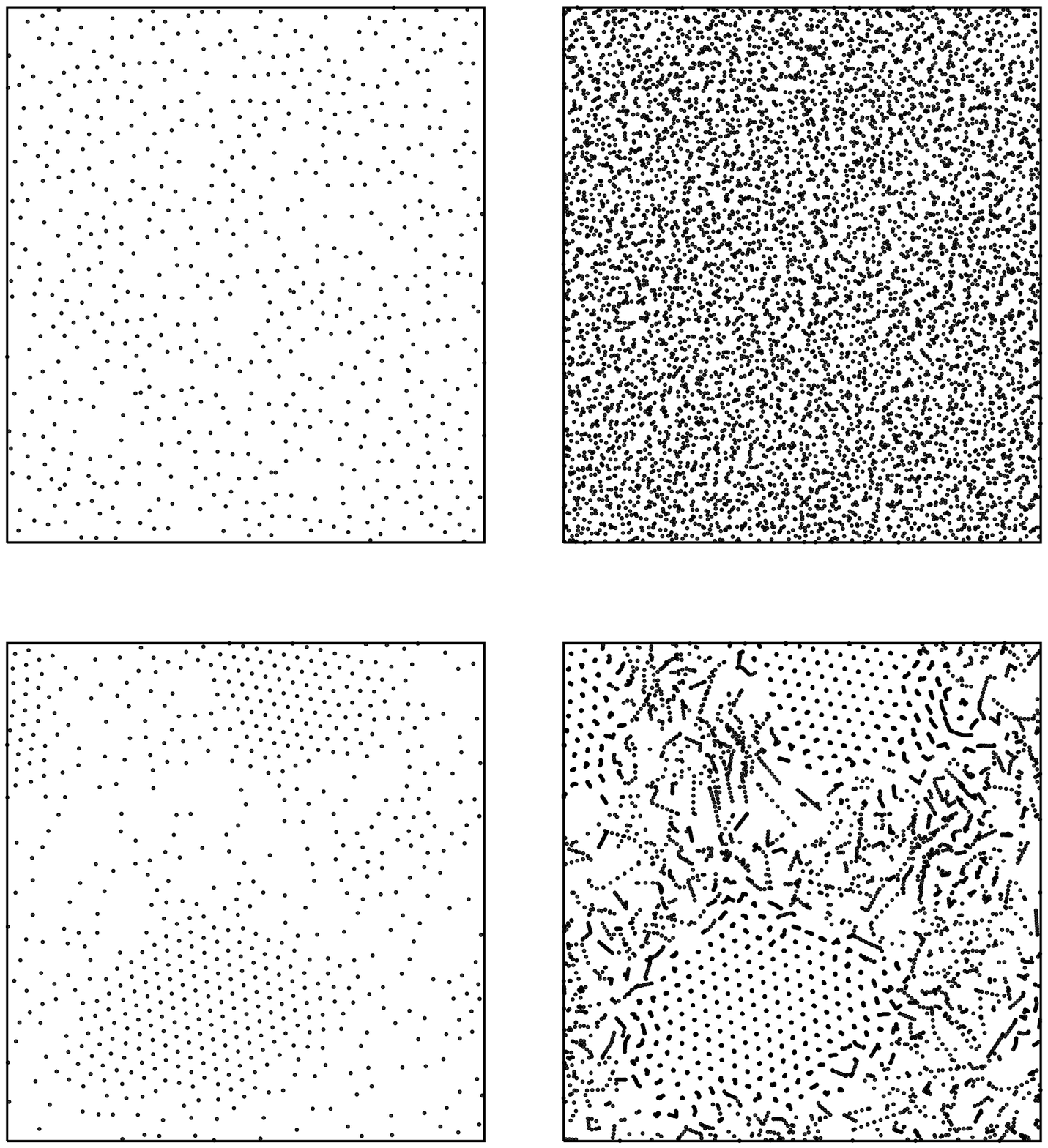}} {\small Fig.~1.  The gas
versus the cluster-gas phase.  Shown are the instantaneous particle
positions (left) and the cumulative positions over 100 consecutive
oscillation cycles (right) for vibration intensities $\Gamma=1.0$ (up)
and $\Gamma=0.6$ (down).}  
\end{figure}

Qualitatively, we observe that above a critical vibration intensity,
$\Gamma>\Gamma_c$, particles are in a gas phase, in which their motion
is random, as seen in Fig.~1. When $\Gamma<\Gamma_c$ in addition to
particles in the gas phase, hexagonal ordered clusters form, as shown
in a snapshot of the system.  Furthermore, particles inside these
hexagonal clusters are stationary, while particles outside the
clusters move appreciably.  This behavior is rather robust as it is
independent of many of the underlying parameters including the
dissipation parameters. Nevertheless, the simulations indicate that
the critical acceleration $\Gamma_c$ is primarily determined by
$\gamma_n$, while the stability of the clusters is governed by
$\gamma_s$. Additionally, the relaxation time scale $\tau$ had to be
sufficiently small for the system to be able to fully relax any
transient behavior.

\begin{figure}
\centerline{\epsfxsize=8cm \epsfbox{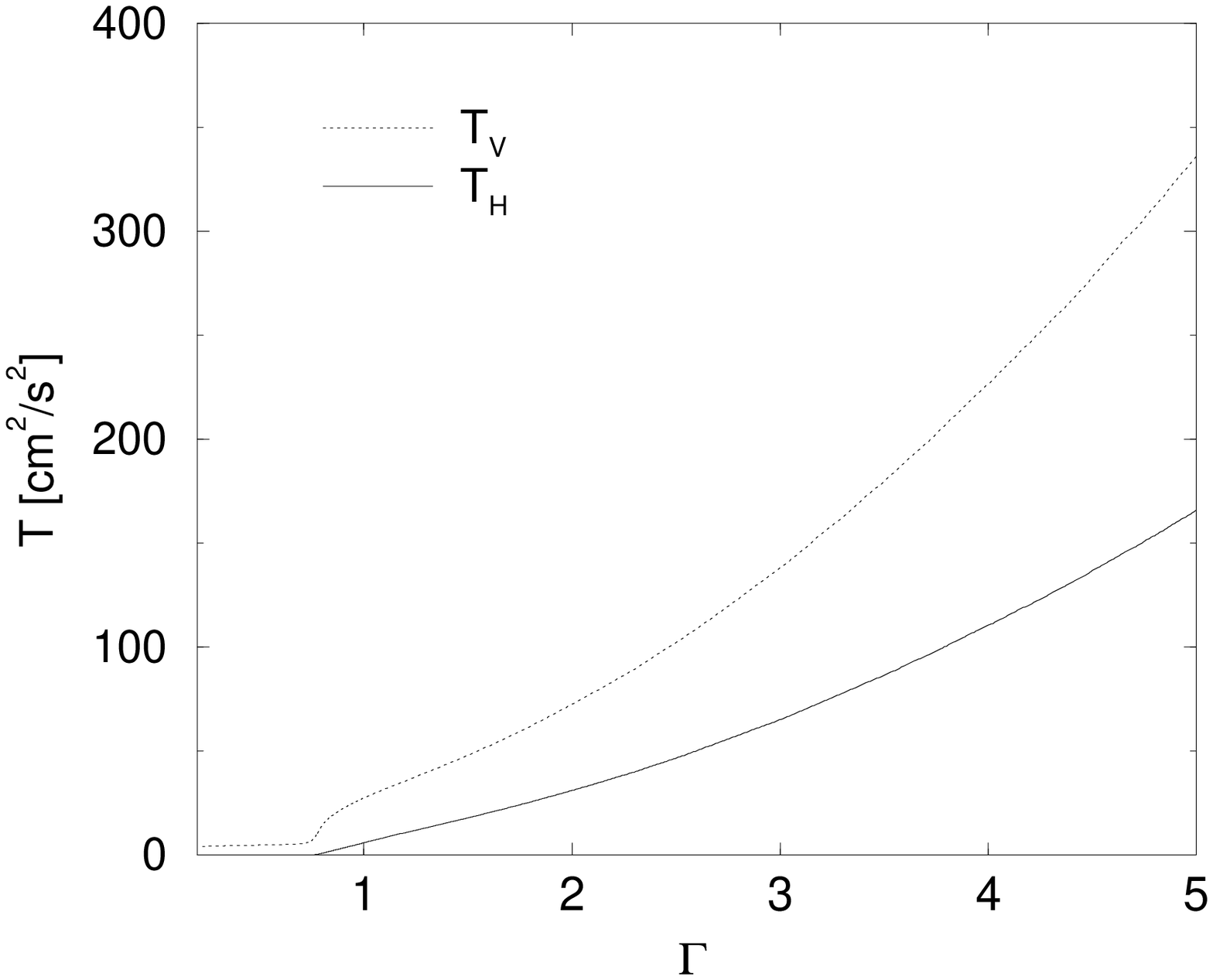}} 
{\small {\bf Fig.~2} The vertical and horizontal temperatures versus the 
vibration strength.}
\end{figure}

Experimental measurements of the horizontal temperature, defined by
$T_H = \langle v_x^2+v_y^2\rangle$, indicate a linear dependence in
the vicinity of the critical point \cite{olafsen2} 
\begin{equation}
 T_H \propto (\Gamma-\Gamma_c).
\end{equation}
Our simulations confirm this linear behavior, as shown in Fig.~2 and
Fig.~3. Furthermore, the horizontal energy practically vanishes below
the transition point, as this quantity decreases by 3 orders of
magnitude for $\Gamma<\Gamma_c$. This is reminiscent of a sharp phase
transition and it is therefore sensible to view $\Gamma_c$ as a
critical point. This linear behavior can be used to estimate the
critical point, and a linear least-square-fit yields $\Gamma_c=0.763$,
in good agreement with the experimental observation, $\Gamma_c=0.77$
\cite{olafsen1}. We conclude that the near critical behavior observed
numerically agrees both qualitatively and quantitatively with the
experimental observations.

Simulations also allow measurements of the vertical velocities. We
find that the vertical energy $T_V=2 \langle{v_z}^2\rangle$ decreases
sharply near the transition point as well.  However, in contrast with
the horizontal energy, it does not vanish below the transition point.
Therefore, the velocities develop a strong anisotropy as the vertical
and the horizontal velocities behave quite differently. This reflects
the fact that the system is far from equilibrium.

\begin{figure}
\centerline{\epsfxsize=8cm \epsfbox{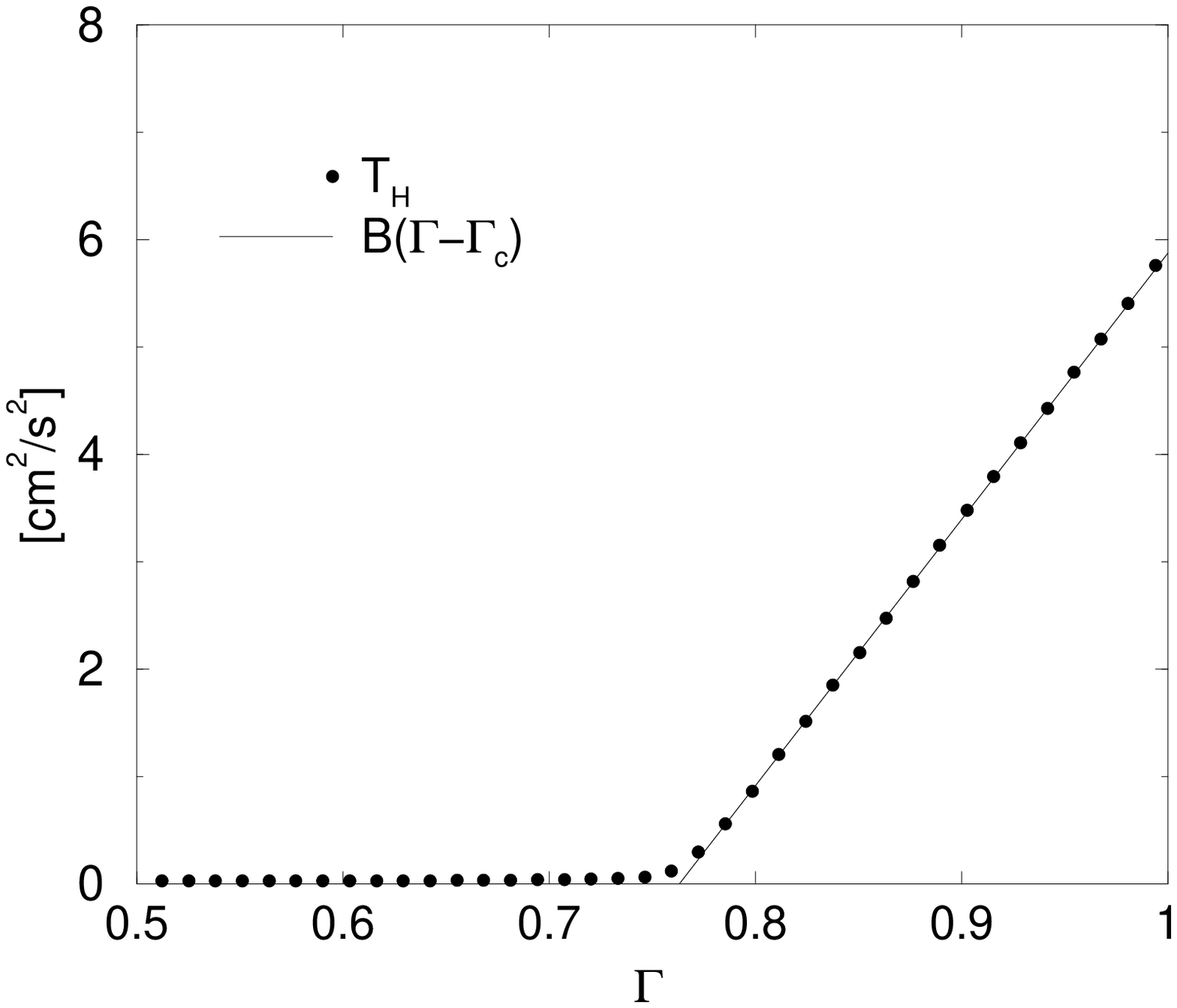}} {\small {\bf Fig.~3}. 
Near critical behavior of the horizontal temperature.  The 
critical acceleration was determined to be $\Gamma_c=0.763$ from 
the linear fit $T_H=B(\Gamma-\Gamma_c)$, plotted as a solid line.}
\end{figure}

More detailed velocity statistics is provided by the velocity
distribution. We observe that at high accelerations the particle
motion is nearly isotropic, i.e., the ratio of horizontal to vertical
energy is of the order unity. Indeed, this ratio approaches a value of
roughly 0.5 (see Fig.~2). In addition, the velocity distribution is
Gaussian (see Fig.~4) and the system is practically three-dimensional.
However, as the acceleration is decreased, the two-dimensional
geometry becomes more and more pronounced. The vertical motion
dominates over the horizontal one, and the horizontal velocity
distribution departs strongly from a Gaussian distribution.  Near the
phase transition point the large velocity tail becomes nearly
exponential (see Fig.~4).  Below the transition point, a significant
fraction of the particles have a nearly vanishing horizontal velocity,
and the distribution of horizontal velocities is strongly enhanced
near the origin.

The deviation from the Gaussian behavior can be quantified using the
kurtosis, defined via the fourth and second moments of the
distribution, $\kappa=\langle v^4\rangle/\langle v^2\rangle^2$.
Indeed, in the limit of high vibration intensities, $\Gamma\gg 1$,
this parameter approaches the Gaussian value $\kappa\to 3$.  On the
other hand, near the phase transition point, i.e, as
$\Gamma\to\Gamma_c$, this parameter approaches the exponential value
$\kappa\to 6$.  It proves useful to examine how the kurtosis depends
on $T_H/T_V$, the ratio between the horizontal and the vertical
energies. As shown in the inset to Fig.~4, the smaller the ratio (or
equivalently, the larger the anisotropy), the larger the deviation
from a Gaussian distribution. Hence, whether the velocity distribution
is Gaussian or not reflects the degree of anisotropy in the particle
motion.  Non-Gaussian distributions has been observed
experimentally\cite{olafsen1,olafsen2,losert2}, theoretically
\cite{godhirsch,puglisi,javier}, and numerically
\cite{gzb,javier,isobe,taguchi} in two-and three-dimensional
geometries.

\begin{figure}
  \centerline{\epsfxsize=8cm \epsfbox{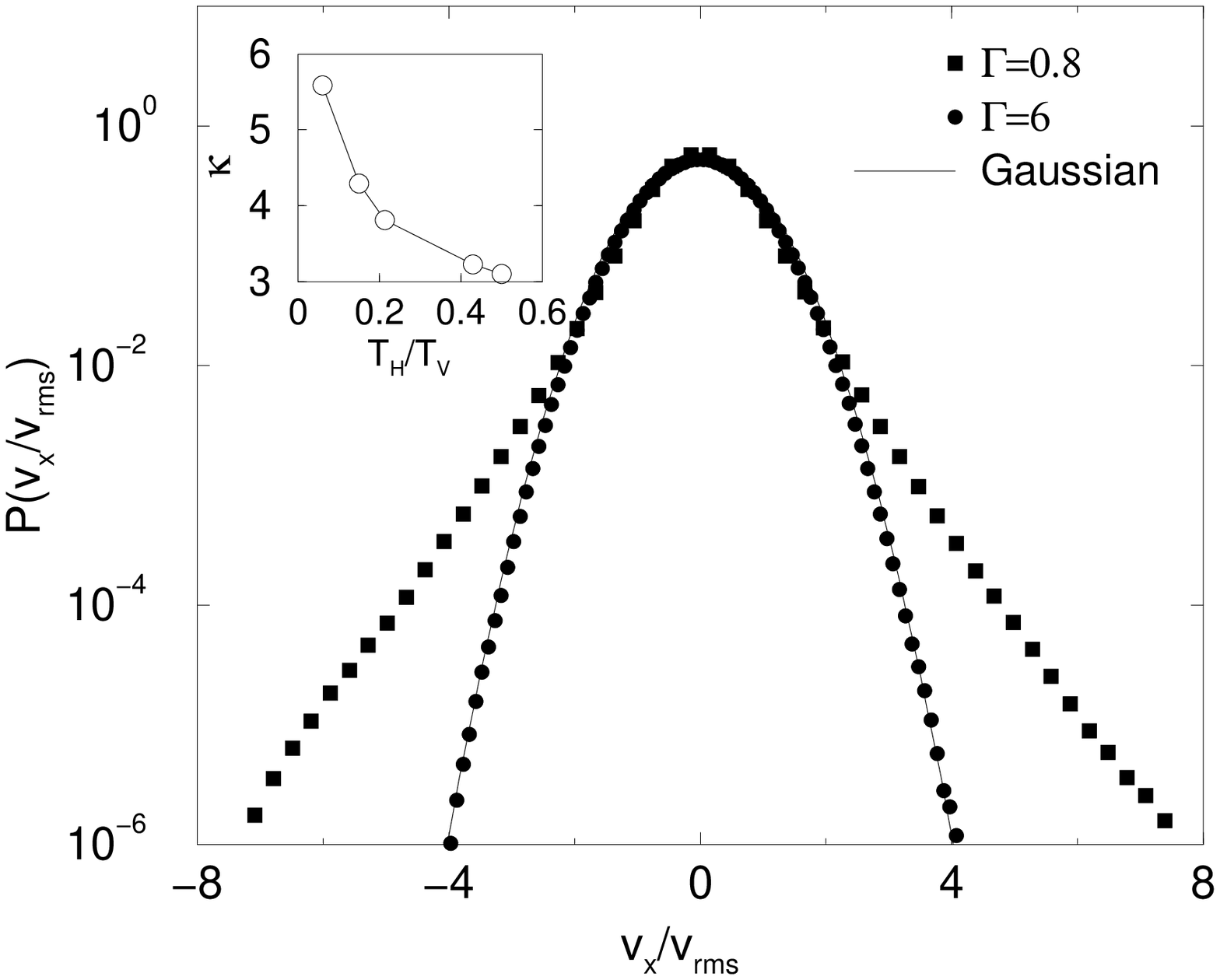}} {\small {\bf Fig.~4}
    The distribution of horizontal velocities in the gas state, and in
    the vicinity of the critical point. The velocities were normalized
    by the RMS velocity $v_{\rm rms}=\langle v_x^2\rangle^{1/2}$. A Gaussian
    distribution is plotted as a reference. The inset plots $\kappa$
    the kurtosis of the distribution of horizontal velocities versus
    $T_H/T_V$ the ratio of horizontal to vertical energies. The two
    extremal points correspond to the data sets plotted in this
    figure, i.e, $\Gamma=0.8$ and $\Gamma=6$.} 
\end{figure}

\begin{figure}
  \centerline{\epsfxsize=8cm \epsfbox{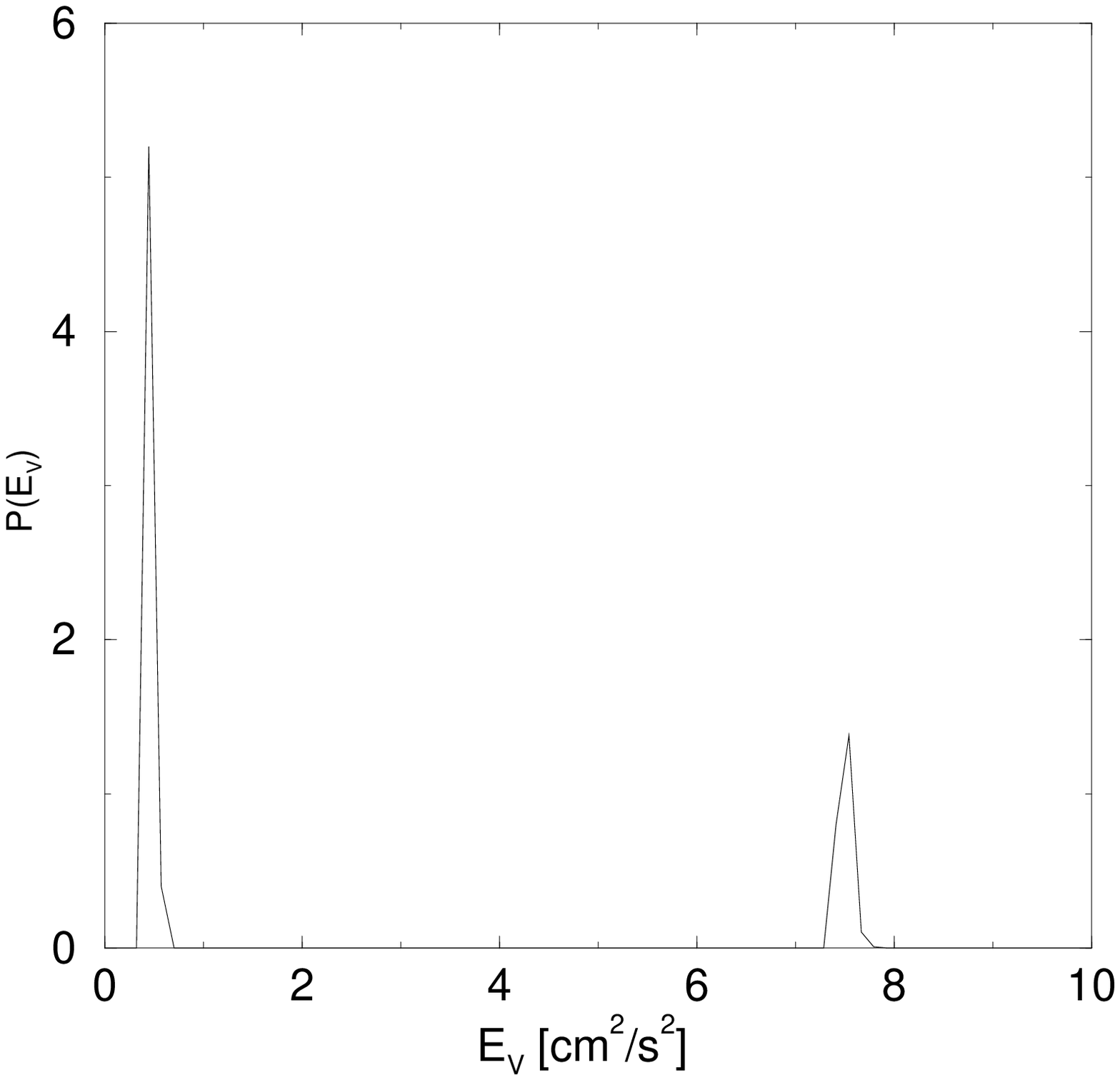}} {\small {\bf Fig.~5}
    The distribution of vertical velocities in the cluster-gas phase.}
\end{figure}

The distribution of vertical velocities can be used to distinguish
between cluster and gas particles.  Indeed, the vertical velocity
distribution changes its character from a unimodal to a bimodal
distribution in the cluster-gas phase (see Fig.~5) with the low
velocity peak corresponding to the cluster particles, and the high
velocity peak corresponding to the gas particles. Interestingly, the
location of the high velocity peak does not change as $\Gamma$
decreases. In fact, this energy can be understood by considering the
first excited state energy of a single particle bouncing on a
vibrating plate, which can be calculated to be \cite{losert1}
\begin{equation}
E_1 = {1\over 6}\left({\pi g \over\omega}\right)^2,  
\end{equation}
or in our case $E_1=8.16\ cm^2/s^2$.  We verified this result by
simulating the motion of a single particle on a vibrating plate.
Interestingly, the energy of the gas particles falls within less then
10\% of this value.  Therefore, below the phase transition point
particles residing in clusters are in the ground state, i.e., they are
moving with the plate.  Furthermore, the rest of the particles
constituting the gas phase are in the first excited state of an
isolated ball on a vibrating surface.  This indicates that particles
in the gas phase are essentially noninteracting.

The vertical velocity distribution can be used to study the fraction
of particles in each phase by simply integrating the area under the
respective energy peaks.  As shown in Fig.~6, $P_0$, the fraction of
particles in the gas phase, is almost independent of the vibration
intensity below the transition point. As the transition point is
approached this fraction rapidly decreases and ultimately vanishes for
$\Gamma\gg \Gamma_c$.  Although this quantity does not undergo a sharp
transition, its behavior is consistent with our previous estimate of
the transition point from the horizontal energy behavior.

\begin{figure}
\centerline{\epsfxsize=8cm \epsfbox{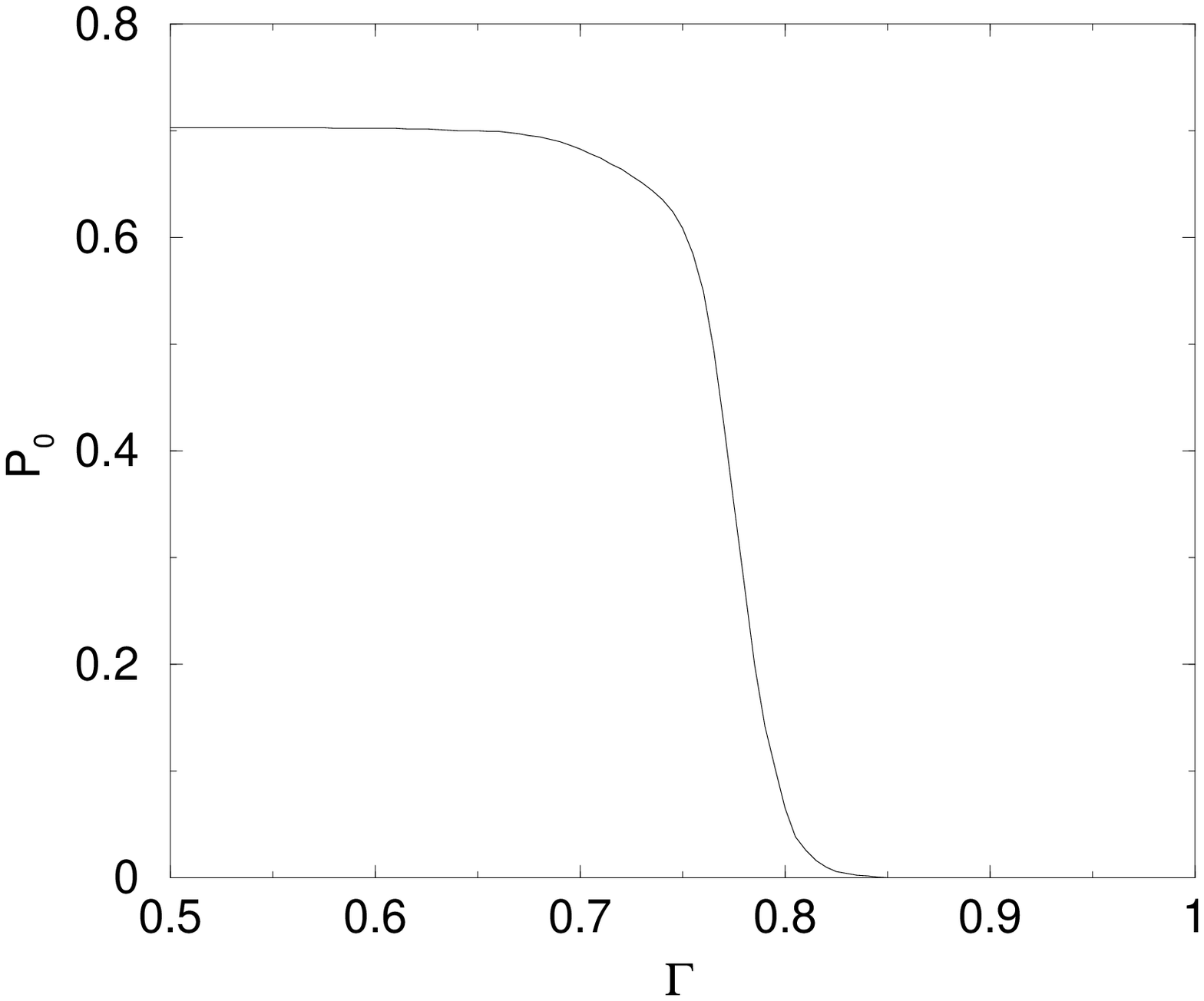}} 
{\small {\bf Fig.~6} The fraction of particles moving with the 
plate. The fraction $P_0$ was obtained from the horizontal velocity 
distribution by integrating the area enclosed under the low velocity 
peak (see Fig.~5).} 
\end{figure}

In summary, we have studied the dynamics of vibrated granular
monolayers using molecular dynamics simulations.  We find that the
transition between the gas and the cluster-gas phases can be regarded
as a sharp phase transition, and that the horizontal energy decreases
linearly near the transition point. We have shown that at high
vibration strengths, the particle motion is isotropic, and the
velocity distributions are Gaussian. The deviation from a Gaussian
distribution were found to be closely related to the degree of
anisotropy in the motion. We have also shown that below the phase
transition, the velocity distribution is bimodal. The cluster
particles move with the plate, while the gas particles behave as a
noninteracting gas, as their energy agrees with the first excited
state of an isolated vibrated particle.

Our results agree both qualitatively and quantitatively with the
experimental data. This shows that the underlying phenomena can be
explained solely by the simulated interactions, i.e., contact force
interactions, no attractive forces, and dissipative collisions.  Other
mechanisms, possibly present in the experiment, such as electrostatic
forces, etc. are therefore not responsible for the phase transition.
It will be interesting to use molecular dynamics simulations to
determine the full phase diagram of this system by varying the density   
and the driving frequency.

\end{multicols}
\end{document}